\documentclass{article}
\usepackage{amsfonts,bezier}
\usepackage{graphics}

\newcommand{\bls}[1]{\renewcommand{\baselinestretch}{#1}}
\def\noi{\noindent}

\makeatletter
\renewcommand{\section}{\@startsection{section}{1}{0pt}%
        {-3.5ex plus -1ex minus -.2ex}{2.3ex plus .2ex}%
        {\large\bf\protect\raggedright}}

\renewcommand{\subsection}{\@startsection{subsection}{2}{0pt}%
        {-3ex plus -1ex minus -.2ex}{1.4ex plus .2ex}%
        {\normalsize\bf\protect\raggedright}}

\renewcommand{\thesubsubsection}%
        {\arabic{section}.\arabic{subsection}.\arabic{subsubsection}.}
\newcommand{\para}{\@startsection{paragraph}{4}{0pt}%
        {1.5ex plus -.5ex minus -.2ex}{-1em}{\normalsize\bf}}

\renewcommand{\@oddhead}{\raisebox{0pt}[\headheight][0pt]{%
   \vbox{\hbox to\textwidth{\rightmark \hfil \rm \thepage
\strut}\hrule}}}
\renewcommand{\@evenhead}{\raisebox{0pt}[\headheight][0pt]{%
   \vbox{\hbox to\textwidth{\thepage \hfil \leftmark \strut}\hrule}}}

\newcommand{\heads}[2]{\markboth{\protect\small\it
#1}{\protect\small\it #2}}

\makeatother

\newcommand{\Title}[1]{\noi {\Large #1} \\}
\newcommand{\Author}[2]{\noi{\large\bf #1}\\[2ex]\noindent{\it #2}\\}

\newcommand{\Abstract}[1]{\vskip 2mm \begin{center}
        \parbox{16.4cm}{\small\noi #1} \end{center}\medskip}

\newcommand{\Ref}[1]{Ref.\,\cite{#1}}
\newcommand{\sect}[1]{Sec.\,#1}

\def\nqq{\hspace*{-2em}}
\def\nhq{\hspace*{-0.5em}}

\def\cm{\hspace*{1cm}}

\def\para{\paragraph}

\def\eq{Eq.\,}
\def\eqs{Eqs.\,}
\def\beq{\begin{equation}}
\def\eeq{\end{equation}}
\def\bear{\begin{eqnarray}}
\def\al{&\nhq}
\def\lal{&&\nqq {}}               
\def\bearr{\begin{eqnarray} \lal}
\def\ear{\end{eqnarray}}
\def\earn{\nonumber \ear}

\def\nn{\nonumber\\ {}}

\def\eql{\al =\al}

\def\e{{\,\rm e}}
\def\d{\partial}

\def\const{{\rm const}}

\def\Jl#1#2{{\it #1\/} {\bf #2},\ }

\def\CQG#1 {\Jl{Clas. Qu. Grav.}{#1}}
\def\DAN#1 {\Jl{Dokl. AN SSSR}{#1}}
\def\GC#1 {\Jl{Grav. \& Cosmol.}{#1}}
\def\GRG#1 {\Jl{Gen. Rel. Grav.}{#1}}
\def\JETF#1 {\Jl{Zh. Eksp. Teor. Fiz.}{#1}}
\def\JMP#1 {\Jl{J. Math. Phys.}{#1}}
\def\NP#1 {\Jl{Nucl. Phys.}{#1}}
\def\PLA#1 {\Jl{Phys. Lett.}{#1A}}
\def\PLB#1 {\Jl{Phys. Lett.}{#1B}}
\def\PRD#1 {\Jl{Phys. Rev.}{D\ #1}}
\def\PRL#1 {\Jl{Phys. Rev. Lett.}{#1}}

\def\GR{general relativity}
\def\sph{spherically symmetric}
\def\ssph{static, spherically symmetric}
\def\fig{Fig.\,}

\def\R{{\mathbb R}}
\def\S{{\mathbb S}}

\heads{K.A. Bronnikov and B. E. Meierovich}
{Multidimensional Global Monopole and Nonsingular Cosmology}

\bls{0.97}
\bls{0.95}
\begin{document}
\thispagestyle{empty}
\rightline{\bf gr-qc/0301084}
\bigskip

\Title
{\bf Multidimensional Global Monopole and Nonsingular Cosmology}

\Author{Kirill A. Bronnikov} {Centre for Gravitation and Fundam.
Metrology, VNIIMS,
        3-1 M. Ulyanovoy St., Moscow 117313, Russia;\\
Institute of Gravitation and Cosmology, PFUR,
        6 Miklukho-Maklaya St., Moscow 117198, Russia\\}

\Author{Boris E. Meierovich} {Kapitza Institute for Physical
Problems, 2 Kosygina St., Moscow 117334, Russia\\ Email:
meierovich@yahoo.com; http://geocities.com/meierovich/}


\Abstract {We consider a spherically symmetric global monopole in
general relativity in $(D=d+2)$-dimensional spacetime. The
monopole is shown to be asymptotically flat up to a solid angle
defect in case $\gamma < d-1$, where $\gamma$ is a parameter
characterizing the gravitational field strength. In the range $
d-1< \gamma < 2d(d+1)/(d+2)$ the monopole space-time contains a
cosmological horizon. Outside the horizon the metric corresponds
to a cosmological model of Kantowski-Sachs type, where spatial
sections have the topology ${\R\times \S}^d$. In the important
case when the horizon is far from the monopole core, the temporal
evolution of the Kantowski-Sachs metric is described analytically.
The Kantowski-Sachs space-time contains a subspace with a
$(d+1)$-dimensional Friedmann-Robertson-Walker metric, and its
possible cosmological application is discussed. Some numerical
estimations in case $d=3$ are made showing that this class of
nonsingular cosmologies can be viable. Other results, generalizing
those known in the 4-dimensional space-time, are derived, in
particular, the existence of a large class of singular solutions
with multiple zeros of the Higgs field magnitude. }

\section{Introduction}

In our recent paper with E. Podolyak \cite{BMP} we considered the
general properties of global monopole solutions in general
relativity, developing some earlier results (see \cite{vilshel,
lieb} and references therein). It was confirmed, in particular,
that the properties of these objects are governed by a single
parameter $\gamma$, squared energy of spontaneous symmetry
breaking in Planck units. For $0 < \gamma < 1$, solutions with
entirely positive (or entirely negative) Higgs field are globally
regular and asymptotically flat up to a solid angle deficit. In
the range $1 < \gamma < 3$, the space-time of the solutions
remains globally regular but contains a cosmological horizon at a
finite distance from the center. Outside the horizon the geometry
corresponds to homogeneous anisotropic cosmological models of
Kantowski-Sachs type, whose spatial sections have the topology
${\R\times \S}^2$. The nonzero symmetry-breaking potential can be
interpreted as a time-dependent cosmological constant, a kind of
hidden vacuum matter. The potential tends to zero at late times,
and the ``hidden vacuum matter'' disappears. This solution with a
nonsingular static core and a cosmological metric outside the
horizon drastically differs from the standard Big Bang models and
conforms to the ideas advocated by Gliner and Dymnikova
\cite{Gliner} that the standard Big Bang cosmology could be
replaced by a globally regular model. A possibility of a
nonsingular isotropic cosmological model had been discussed by
Starobinsky \cite{Starobinsky}.

The lack of isotropization at late times did not allow us to directly apply
the toy model of a global monopole to the early phase of our Universe.
Though, this circumstance does not seem to be a fatal shortcoming of the
model since the anisotropy of the very early Universe could be damped later
by particle creation, and the further stages with low energy densities might
conform to the standard isotropic Friedmann cosmology. Another idea is to
add a comparatively small positive quantity $\Lambda $ to the
symmetry-breaking potential (to ``slightly raise the Mexican hat''). It can
change nothing but the late-time asymptotic which will be de Sitter,
corresponding to the added cosmological constant $\Lambda$. These ideas
deserve a further study.

In this paper we study the gravitational properties of global monopoles in
multidimensional general relativity. Such considerations can be of interest
in view of numerous attempts to construct a unified theory using the ideas
of supersymmetry in high dimensions. Objects like multidimensional
monopoles, strings and other topological defects might form due to phase
transitions in the early Universe at possible stages when the present three
spatial dimensions were not yet separated from others, and a greater number
of dimensions were equally important.

More specifically, we consider a self-gravitating hedgehog-type
configuration of a multiplet of scalar fields with the Mexican-hat potential
$V= (\lambda/4)(\phi^2 - \eta^2)^2$ in a $D$-dimensional space-time with the
structure $\R_t \times \R_\rho \times \S^d$ ($d=D-2$), where $\R_\rho$ is
the range of the radial coordinate $\rho$ and $\R_t$ is the time axis. The
properties of such objects generalize in a natural way the results obtained
in \Ref {BMP} and earlier papers (e.g., \cite{vilshel, lieb}). Thus, for
small values of the parameter $\gamma = 8\pi G \eta^2$ characterizing the
gravitational field strength, the solutions are asymptotically flat up to a
solid angle deficit. Within a certain range $d-1 < \gamma <
\overline{\gamma}(d)$ the solutions are nonsingular but contain a Killing
horizon and a cosmological metric of Kantowski-Sachs type outside it. In the
important case when the horizon is far from the monopole core, the temporal
evolution of the Kantowski-Sachs metric is described analytically.
The upper bound $\overline{\gamma}(d)$, beyond which there are no static
solutions with a regular center, is also found analytically.

The above description concerned solutions with totally positive
(or totally negative) scalar field magnitude $\phi$. As in
\cite{BMP}, we here also find a class of solutions with any number
$n$ of zeros of $\phi(r)$, existing for $\gamma < \gamma_n(d)$,
where the upper bounds $\gamma_n$ are found analytically. All
solutions with $n > 0$ describe space-times with a regular center,
a horizon and a singularity beyond this horizon.

We also discuss a possible cosmological application of multidimensional
global monopoles, which can be of particular interest in the case of
5-dimensional space-time with 3-dimensional spheres $\S^d$. In this case the
Kantowski-Sachs type model outside the horizon has the spatial topology
$\R\times \S^{3}$. It is anisotropic in 4 dimensions but the 3-dimensional
spheres $\S^3$ are isotropic. The anisotropy is thus related only to the
fourth coordinate $t$, which is spatial outside the horizon and is a cyclic
variable from the dynamical viewpoint. If we identify $\S^3$ with the
observed space, ignoring the extra coordinate, we arrive at a closed
cosmological model, with the Friedman-Robertson-Walker line element in the
ordinary (3+1)-dimensional space-time.

A natural question arises: why is the fourth spatial dimension
unobservable today? An answer cannot be found within our
macroscopic theory without specifying the physical nature of
vacuum. The conventional Kaluza-Klein compactification of the
extra dimension on a small circle is not satisfactory in our case
since it leads to a singularity at the horizon (as will be
demonstarted in \sect 3). So we leave this question open and note
that the global monopole model has a chance to describe only the
earliest phase of the cosmological evolution. Its later stages
should involve creation of matter and a sequence of phase
transitions possibly resulting in localization of particles across
the $t$ direction. We then obtain a model with a large but
unobservable extra dimension, similar in spirit to the widely
discussed brane world models, see the reviews \cite{Rubakov,
Maart, Lang} and references therein.

The solutions of interest appear when the symmetry breaking scale $\eta$ is
large enough, and one may suspect that quantum gravity effects are already
important at this energy scale. We show in \sect 2.3 that this is not the
case when the monopole core radius is much greater than the Planck length:
the curvature and energy scales are then in the whole space much smaller
than their Planckian values.

The existence of nonsingular models of the early Universe on the
basis of classical gravity supports the opinion that our Universe
had never undergone a stage described by full quantum gravity.
Apart from those discussed here, such models are now rather
numerous (\cite{BMP,Gliner,dym,brfab}, see also refrences
therein). All of them are evidently free of the long-standing
problems of the standard Big Bang cosmology connected with the
existence of multiple causally disconnected regions \cite{MTW,
Sahni and Starobinsky}.

The paper is organized as follows. In \sect 2 we analyze the properties of
a global monopole in $D=d+2$ dimensions (one time coordinate and $d+1$
spatial coordinates). It is a generalization of our previous results
\cite{BMP}. The particular case $d=3$ is studied in more detail in \sect 3
along with its possible cosmological application. Unless otherwise
indicated, we are using the natural units $\hbar = c =1$.

\section{Multidimensional global monopole}

\subsection{General characteristics}

The most general form of a \ssph\ metric in $D=d+2$ dimensions is
\beq
    ds^2=\e^{2F_0}dt^2-\e^{2F_1}d\rho ^2-\e^{2F_{\Omega}}d\Omega ^2.
    \label{metric general}
\eeq
Here $d\Omega^2 = d\Omega_d^2$ is a linear element on a $d$-dimensional
unit sphere, parametrized by the angles $\varphi_1, \ldots, \varphi_d$:
\[
    d\Omega_d^2=d\varphi _{d}^2+\sin ^2\varphi
    _{d}\left( d\varphi _{d-1}^2+\sin ^2\varphi _{d-1}\left(
    d\varphi _{d-2}^2+...+\sin ^2\varphi _{3}\left( d\varphi
    _2^2+\sin ^2\varphi _2d\varphi _1^2\right) ...\right)
    \right);
\]
$F_0$, $F_1$ and $F_{\Omega }$ are functions of the radial coordinate
$\rho $\ which is not yet specified. The nonzero components of the
Ricci
tensor are (the prime denotes $d/d\rho $)
\bear
R_0^0 \eql \e^{-2F_1}\left[
    F_0^{\prime \prime }+F_0^{\prime }\left( F_0^{\prime
    }+dF_{\Omega }^{\prime }-F_1^{\prime }\right) \right],
\nn
   R_{\rho }^{\rho } \eql \e^{-2F_1}\left[ dF_{\Omega }^{\prime \prime
    }+F_0^{\prime \prime }+dF_{\Omega }^{\prime 2}+F_0^{\prime
    2}-F_1^{\prime }\left( F_0^{\prime }+dF_{\Omega }^{\prime
    }\right) \right],   \label{Ricci tensor}
\nn
 R_2^2 \eql ...=R_{d+1}^{d+1}=-\left( d-1\right) \e^{-2F_{\Omega
    }}+\e^{-2F_1} \left[ F_{\Omega }^{\prime \prime }+F_{\Omega
    }^{\prime }\left( F_0^{\prime }+dF_{\Omega }^{\prime
    }-F_1^{\prime }\right) \right]
\ear

A global monopole with a nonzero topological charge can be constructed
with a multiplet of real scalar fields $\phi ^{a}$ $\left(
a=1,2,...,d+1\right)$
comprising a hedgehog configuration in $d+1$\ spacial dimensions%
\footnote{A 7D universe with a global monopole with a hedgehog
    configuration of scalar fields only in three extra dimensions was
    recently considered by Benson and Cho \cite{Benson Cho}. Our approach is
    different. We consider a hedgehog configuration in all $D-1$ space
    dimensions of the $D$-dimensional spacetime.}:
\[
    \phi ^{a}=\phi \left( \rho \right) n^{a}\left( \varphi
        _1,...,\varphi _{d}\right),
\]
where $n^{a}( \varphi _1,...,\varphi _{d}) $ is a unit vector
($n^a\, n^a = 1$) in the $(d+1)$-dimensional Euclidean target space,
with the components
\bear
        n^{d+1} \eql \cos \varphi _{d},
\nn
    n^{d} \eql \sin \varphi _{d}\cos \varphi _{d-1},
\nn
    n^{d-1}\eql \sin \varphi _{d}\sin \varphi _{d-1}\cos \varphi _{d-2},
\nn
    && \cdots
\nn
    n^{d-k} \eql \sin \varphi _{d}\sin \varphi _{d-1}...\sin \varphi
        _{d-k}\cos \varphi _{d-k-1},
\nn
    && \cdots
\nn
    n^2 \eql \sin \varphi_{d}...\sin \varphi _2\cos \varphi_1,
\nn
    n^1 \eql \sin \varphi _{d}...\sin \varphi _2\sin \varphi_1.
\earn

The Lagrangian of a multidimensional global monopole in general
relativity has the form
\[
    L=\frac{1}{2}\d _{\mu }\phi
    ^{a}\d ^{\mu }\phi ^{a}-V (\phi) + \frac{R}{16\pi G},
\]
where $R$ is the scalar curvature, $G$ is the $D$-dimensional
gravitational constant, and $V (\phi) $ is a symmetry-breaking potential
depending on $\phi =\pm \sqrt{\phi ^{a}\phi ^{a}}$, and it is natural
to choose  $V$ as the Mexican-hat potential
\beq
    V= \frac{\lambda}{4} (\phi^2 - \eta^2)^2
    = \frac{\lambda \eta ^{4}}{4}( f^2-1)^2.     \label{mexhat}
\eeq
We have introduced the normalized field magnitude $f=\phi (\rho)/\eta$
playing the role of the order parameter. The model has a global $SO(d+1)$
symmetry, which can be spontaneously broken to $SO(d)$; $ \eta^{2/d} $ is
the energy of symmetry breaking.
We are using natural units, so that $\hbar =c=1$.

The Einstein equations can be written as
\beq                                                       \label{EE}
    R_{\mu }^{\nu }=-8\pi G\widetilde{T}_{\mu }^{\nu }=-8\pi G\left(
        T_{\mu }^{\nu }-\frac{1}{d}T\delta _{\mu }^{\nu }\right) ,
\eeq
where\ $T_{\mu }^{\nu }$ is the energy-momentum tensor. The
nonzero components of $\widetilde{T}_{\mu }^{\nu }$ are
\bear
    \widetilde{T}_0^0 \eql -\frac{2}{d}V,
\nn
    \widetilde{T}_{\rho}^{\rho } \eql -\e^{-2F_1}f^{\prime
    2}-\frac{2}{d}V,
\nn
    \widetilde{T}_2^2
   \eql...=\widetilde{T}_{d+1}^{d+1}=-\e^{-2F_\Omega} f^2-\frac{2}{d}V.
\earn

Let us use the quasiglobal coordinate $\rho$ specified by the condition
\[
    F_0 + F_1=0,
\]
which is a convenient gauge for \sph\ systems with Killing horizons.
Introducing the functions $A(\rho) = \e^{2F_0}=\e^{-2F_1}$ and $r(\rho)
=\e^{F_{\Omega }}$, we reduce the metric to the form
\beq
    ds^2=A( \rho ) dt^2-\frac{d\rho ^2}{A( \rho)}
    - r^2(\rho) d\Omega ^2,                 \label{metric}
\eeq
and get the following equations:
\bear                                                   \label{e-phi}
    \left( Ar^{d}\phi'\right)' - d r^{d-2}\phi \eql r^{d}\frac{\d
        V}{\d \phi },
\\
    r^{\prime \prime}\eql -\frac{8\pi G}{d}r\phi ^{\prime 2}, \label{01}
\\
    \left(r^{d}A' \right)' \eql -\frac{32\pi G}{d}r^{d}V,        \label{00}
\\
    A\left( r^2\right)''-r^2A'' - \left( d-2\right) r^{3}r'\left(
    \frac{A}{r^2}\right)' \eql 2( d-1-8\pi G\phi^2)          \label{02}
\ear
for the unknown functions $\phi ( \rho ) $, $A(\rho)$ and $r( \rho)$.
Only three of these four equations are independent: the scalar field
equation (\ref{e-phi}) follows from the Einstein equations
(\ref{01})--(\ref{02}) due to the Bianchi identities.

\eqs (\ref{e-phi})--(\ref{00}) have the same structure as
\eqs (13)--(15) in \cite{BMP}. General properties of \eqs
(\ref{e-phi})--(\ref{00}) with arbitrary value of $d$\
are the same as for $d=2$, and the classification of their solutions is
also the same. In particular: if $V( \phi ) >0$, the system with a regular
center can have either no horizon or one simple horizon; in the latter case,
its global structure is the same as that of de Sitter spacetime. Below we
will concentrate our attention on the solutions belonging to class (a1)
according to \cite{BMP}, i.e., $r(\rho)$ is monotonically growing from
zero to infinity as $\rho \to \infty$, while $A(\rho)$ changes from $A=1$ at
the regular center to $A_{\infty} < 0$ as $\rho \to \infty$, with a
cosmological horizon (where $A = 0$) at some $\rho=\rho_h$.

\eq (\ref{02}) is a second-order
linear inhomogeneous differential equation with respect to $A$. The
corresponding homogeneous equation has the evident special solution
$A( \rho) = \const \times r^2 ( \rho )$.
This allows one to express $A( \rho ) $ in terms of $r( \rho) $
and $\phi (\rho) $ in an integral form:
\beq
    A=C_1r^2-C_2r^2\int_{\rho }^{\infty
    }\frac{d\rho _1}{r^{d+2}\left( \rho _1\right)
    }+2r^2\int_{\rho }^{\infty }\frac{d\rho _1}{r^{d+2}\left( \rho
    _1\right) }\int_0^{\rho _1}d\rho _2r^{d-2}\left( \rho
    _2\right) \left[ d-1-8\pi G\phi ^2\left( \rho _2\right)
        \right]     \label{A(ro) general}
\eeq

Consider solutions with a large $r$ asymptotic such that $r(\rho)
\to \infty$ and $r'(\rho) \to \const >0$ as $\rho \to \infty$. \eq
(\ref{01}) gives $r'$ as $\int [r\phi'^2]d\rho$, and its
convergence as $\rho\to \infty$ implies a sufficiently rapid decay
of $\phi'$ at large $\rho$, so that $\phi\to \phi_\infty = \const$
as $\rho\to \infty$. The potential $V$ then tends to a constant
equal to  $V(\phi_\infty)$. Furthermore, \eq (\ref{00}) shows that
$A(\rho)$ can grow at large $r$ at most as $r^2$ and, lastly,
substitution of the asymptotics of $\phi(\rho)$, $A(\rho)$ and
$r(\rho)$ into \eq(\ref{e-phi}) leads to $dV/d\phi \to 0$ as
$\rho\to \infty$. So the condition that there exists a large $r$
asymptotic, applied to the field equations, implies that at this
asymptotic the scalar field tends either to an extremum of the
potential $V (\phi)$, or to an inflection point with zero
derivative. For the Mexican hat potential it can be either the
maximum at $\phi=0$ (the trivial unstable solution for $\phi$ and
de Sitter metric with the cosmological constant $2\pi
G\lambda\eta^4$) or a minimum of $V$, where $f=1$ and $V=0$. For a
``slightly raised Mexican hat'' (the potential (\ref{mexhat}) plus
a small constant $V_+$) we have a de Sitter asymtotic with $f=1$
and $V=V_+$.

A regular center requires that $A = A_c + O(r^2)$ and $A r'^2 \to
1$ as $\rho\to \rho_c$ such that $r(\rho_c) =0$. Without loss of
generality we put $\rho_c =0$ and $A_c = 1$.

For the potential (\ref{mexhat}), regularity at $\rho = 0$ and the
asymptotic condition at $\rho \to \infty $ lead to $C_1=C_2=0$,
and from (\ref{A(ro) general}) we have \beq
    A(\rho) =2r^2(\rho)
    \int_{\rho }^{\infty }\frac{d\rho _1}{r^{d+2}\left( \rho
    _1\right) }\int_0^{\rho _1}d\rho _2r^{d-2}\left( \rho
    _2\right) \left[ d-1-8\pi G\phi ^2\left( \rho _2\right)
        \right] .                   \label{A(ro)1}
\eeq

\eq (\ref{00}) provides another representation for $A(\rho)$
satisfying the regular center conditions:
\beq
    A(\rho) =1-\frac{32\pi
    G}{d}\int_0^{\rho }\frac{d\rho _1}{r^{d}\left( \rho
    _1\right) }\int_0^{\rho _1}d\rho _2r^{d}\left( \rho
        _2\right) V\left( \rho _2\right).            \label{A(ro)2}
\eeq
From (\ref{A(ro)1}) we find the limiting value of $A$ at $\rho \to\infty$:
\beq
    A( \infty) =\frac{d-1-\gamma }{\alpha ^2( d-1) },\cm
        \gamma =8\pi G\eta ^2,                 \label{A(infinity)}
\eeq
where $\alpha =dr/d\rho $ at $\rho \to \infty$:
\[
    \alpha =1-\frac{8\pi G}{d}\int_0^{\infty }r(\rho) \phi
        ^{\prime 2}(\rho) d\rho .
\]

    \eq (\ref{A(infinity)})  shows that
    $\gamma= d-1$ is a critical value of $\gamma$:
    the large $r$ asymptotic can be static only if $\gamma\leq d-1$;
    for $\gamma < d-1$ it is flat up to a solid angle deficit, in full
    similarity to the conventional case $d=2$ \cite{vilshel,BMP}.
    If $\gamma > d-1$, then $A(\infty) <0$, and there is a horizon at
some
    $\rho =\rho_{h}$ where $A =0$. From (\ref{A(ro)2}),
\[
    \frac{32\pi G}{d}\int_0^{\rho _{h}}\frac{d\rho _1}{r^{d}\left(
    \rho _1\right) }\int_0^{\rho _1}d\rho _2r^{d}\left( \rho
        _2\right) V\left( \rho _2\right) =1,
\]
    and so we have
\beq
    A(\rho) =-\frac{32\pi
    G}{d}\int_{\rho _{h}}^{\rho }\frac{d\rho _1}{r^{d}\left( \rho
    _1\right) }\int_0^{\rho _1}d\rho _2r^{d}\left( \rho
    _2\right) V\left( \rho _2\right)  \label{A(ro,rh)}
\eeq
    The $\gamma$ dependence of $\rho_h$, where $\gamma =8\pi G\eta ^2$,
can
    be found from the relation
\beq
    \frac{32\pi G}{d}\int_{\rho _{h}}^{\infty }
    \frac{d\rho _1}{r^{d}( \rho _1) }\int_0^{\rho _1}d\rho
    _2r^{d}\left( \rho _2\right) V( \rho _2) =-\frac{d-1-\gamma}
    {\alpha ^2(d-1) }.          \label{ro h (gamma)}
\eeq

\subsection{Large $r$ asymptotic}

From (\ref{e-phi}) we can find the asymptotic behavior of the field $f(\rho)$
and the potential $V(\rho)$ at $r\to \infty $. Thus, at large $\rho$ we
have $A \to A(\infty)$, see (\ref{A(infinity)}), and the field equation
(\ref{e-phi})\ reduces to
\[
    \frac{1}{r^{d}}\frac{d}{dr}\left( r^{d}\frac{df}{dr}\right)
    -\frac{d-1}{\gamma -d+1}\left[ \lambda \eta ^2\left(
    1-f^2\right) -\frac{d}{r^2}\right] f=0,\cm   r\to\infty .
\]
A regular solution of this equation must tend to unity
as $r\to \infty ,$ and for $\psi =1 - f$ we have the linear equation
\beq
    \psi ,_{rr}+\frac{d}{r}\psi,_{r}
    +\frac{2\lambda \eta ^2( d-1) }{\gamma
    -d+1}\left( \psi -\frac{d}{2\lambda \eta ^2r^2}\right)
        =0,\cm                    r\to \infty .  \label{Psi,rr =}
\eeq
The general solution of the corresponding homogeneous equation
\[
    \psi_0,_{rr}+\frac{d}{r}\psi _0,_{r}+\frac{2\lambda \eta
                ^2(d-1) }{\gamma -d+1}\psi _0=0
\]
can be expressed in terms of Bessel functions:
\[
    \psi _0(r)  = r^{-(d-1)/2}
        \left[C_1 J_{-\frac{d-1}{2}}\biggl(\frac{r}{r_0}\biggr)
    + C_2  Y_{-\frac{d-1}{2}}\biggl(\frac{r}{r_0}\biggr)\right],
\cm
    r_0^2 = \frac{\gamma-d+1}{2\lambda \eta ^2(d-1)}.
\]
A special solution of the inhomogeneous equation
(\ref{Psi,rr =})\ at $r\to \infty $\ is
\[
    \psi =\frac{d}{2\lambda \eta ^2r^2}+O\left(
    \frac{1}{r^{4}}\right) .
\]
The general solution of \eq (\ref{Psi,rr =}) gives the following
asymptotic behavior for the Higgs field magnitude $f$\ as $r\to \infty$:
\beq
    f(r) =1-\frac{d}{2\lambda \eta
    ^2r^2}-\frac{C}{(\lambda \eta^2r^2)^{d/4}}\sin \left(
    \frac{r}{r_0}+\frac{\pi d}{4}+\varphi \right) ,\cm
        r\to \infty .              \label{f asymptotics}
\eeq

\begin{figure}\centering
    \includegraphics{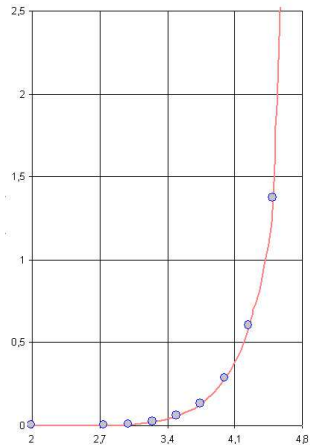}
    \caption{\label{fig:Figure1} The function $C(\gamma)$ found
    numerically for $d=3$}
\end{figure}

Due to the boundary conditions imposed,
the integration constants $C$\ and $\varphi $ are functions of $d$\
and $\gamma$ which can be found numerically. The function
$C(\gamma) $ for $d=3$ is presented in \fig 1. From (\ref{f
asymptotics}) we find the asymptotic behavior of $V$:
\beq
    V(r) =\frac{\lambda \eta^{4}}{4}\left[ \frac{d}{\lambda \eta
    ^2r^2}+\frac{2C}{(\lambda \eta^2r^2)^{d/4}}\sin \left(
    \frac{r}{r_0}+\frac{\pi d}{4}+\varphi \right) \right]^2,
    \qquad r\to \infty .                \label{V(r)}
\eeq

\subsection {Bounds of the classical regime and the monopole core}

    Of certain interest are solutions with cosmological large $r$ behavior,
    i.e., those with $\gamma > d-1$. The latter condition means that the
    scalar field, approaching $\eta$ at large $r$, actually takes near-
    or trans-Planckian values.

    Indeed, in $D$ dimensions, the Planck length $l_D$ and mass $m_D$ are
    expressed in terms of the gravitational constant $G = G_D$ as
\[
    l_D = G_D^{1/d}, \cm m_D = G_D^{-1/d}, \cm  d = D-2.
\]
    Therefore $\eta^2 = \gamma/(8\pi G) = \gamma m_D^{d}/(8\pi)$, and,
    in the case of interest $\gamma \sim d$, we have
\beq
       \eta \sim (m_D)^{d/2}\sqrt{d/(8\pi)}.       \label{eta-pl}
\eeq

    We can, however, remain at sub-Planckian curvature values, thus avoiding
    the necessity to invoke quantum gravity, if we require sub-Planckian
    values of the potential $V$ in the whole space, i.e.,
    $8\pi G V = 2\pi G\,\lambda\eta^4 \ll m_D^2$, whence it follows, for
    $\eta$ given by (\ref{eta-pl}),
\beq
    \lambda \ll \frac{32\pi}{d^2} m_D^{2-d}.          \label{lam-clas}
\eeq
    We can thus preserve the classical regime even with large $\eta$ by
    choosing sufficiently small values of $\lambda$. In terms of lengths,
    this condition is equivalent to the requirement that the monopole core,
    radius $r_{\rm core} = 1/(\sqrt{\lambda}\eta)$ is much greater than
    the Planck length:
\beq
       \frac{1}{\sqrt{\lambda}\eta} \gg l_D.              \label{clas}
\eeq

    One may notice that this condition is external with respect to the theory
    since general relativity does not contain an internal restriction on the
    gravitational field strength. Moreover, in ordinary units, our
    dimensionless gravitational field strength parameter, expressed as
    $\gamma=8\pi Gc^{-4} \eta^2$, does not contain $\hbar$. Only when we
    compare the characteristic length existing in our theory, $r_{\rm
    core}$, with the Planck length $l_D=(\hbar G/c^3)^{1/d}$, we obtain the
    restriction (\ref{lam-clas}) or (\ref{clas}).

    Let us now discuss the solutions for $\gamma$ slightly exceeding the
    critical value $d-1$. In case $\gamma -(d-1) \ll 1$, the horizon
    radius $r_{h}$ is much greater than $r_{\rm core},$ and the constant
    $C$ turns out to be negligibly small (this is confirmed numerically, see
    \fig 1). Then the integrand in the internal integrals in
    (\ref{A(ro)2}), (\ref{A(ro,rh)}), (\ref{ro h (gamma)}) at large
    $\rho_2$ is
\[
    d\rho_2 r^d (\rho_2) V( \rho _2) \approx
            \frac{d^2}{4\alpha \lambda }\frac{dr}{r^{4-d}}.
\]
    The main contribution to the above internal integrals
    comes from the monopole core if $d < 3$ and from the upper limit if
    $d>3$. In case $d=3$ it is a logarithmic integral. As a result, we have
    different behavior%
\footnote
       {This is the only important qualitative
        difference between the general case $d\geq 3$ and the
        particular case $d=2,$ considered in \cite{BMP}.}
    of $\rho _{h}( \gamma) $ at $\gamma -(d-1) \ll 1$ for $d=2$ and
    $d\geq 3$.

    For $d=2$ (4-dimensional \GR),
\[
    \int_0^{\rho _1}d\rho _2r^{d}( \rho_2) V( \rho _2)
    \approx \int_0^{\infty}d\rho_2r^2(\rho_2) V( \rho _2)
    = \const,
\]
    and from (\ref{ro h (gamma)}) we find,
    in agreement with \cite {BMP}, that the horizon radius $r_{h}$\ is
    inverse proportional to $\gamma -1$:
\[
       r_{h}= \const/(\gamma -1),\cm \gamma -1 \ll 1,\cm d=2.
\]
    For $d>3$ we find that at $\gamma -(d-1) \ll 1$ the horizon radius
    $r_{h}$ is inverse proportional to the square root of $\gamma -(d-1)
    \ll 1$:
\beq
     r_{h}=\sqrt{\frac{\gamma d(d-1) }{2( d-3) ( \gamma -d+1)}
    \frac{1}{\lambda \eta ^2}},\cm
     r_{h}^2\gg \frac{1}{\lambda \eta ^2},\qquad d>3.  \label{r h at d>3}
\eeq

    It is thus confirmed that, for $\gamma - (d-1) \ll 1$, the horizon is
    located far from the monopole core, $r_{h}^2\gg 1/(\lambda
    \eta^2)$. Then the function $A(r) $ at $r>r_{h}$ can be found analytically. In
    this case $r(\rho) $ is a linear function at $r > r_{h}$, and
    $dr=\alpha\, d\rho$.  From (\ref{A(ro,rh)}) at $r>r_{h}$ we find
\beq
    A(r) =-\frac{\gamma + 1 - d}{\alpha^2 (d-1)}
    \left( 1-\frac{r_{h}^{d-1}}{r^{d-1}}\right) +\frac{\gamma
    d}{2\alpha ^2( d-3) \lambda \eta ^2r^2}\left[
    1-\left( \frac{r_{h}}{r}\right) ^{d-3}\right]. \label{A(r) at d>3}
\eeq
    The condition of applicability of (\ref{A(r) at d>3}) is $l_D \ll
    r_h. $ In view of $r_{\rm core} \ll r_h $ it is less restrictive than
    the condition (\ref{clas}).

\subsection{Solutions with $f(\phi)$ changing sign}

    As in \Ref {BMP}, numerical integration of the field equations
    shows that, in addition to the solutions with totally positive
    (or totally negative) $f(u)$, there are also solutions with a regular
    center such that $f(u)$
    changes its sign $n$ times where $n\geq 1$. All these solutions
    exist for $\gamma < \gamma_n(d)$, where $\gamma_n(d)$ are some
    critical values of the parameter $\gamma$. In case $n>0$, all of them
    have a horizon, and the absolute value of $f$ at the horizon $
    |f_{h,n}(\rho_h)|$ is a decreasing function of $\gamma$, vanishing as
    $\gamma \to \gamma_n-0$.  Moreover, as $\gamma  \to \gamma_n(d)$, the
    function $f(u)$ vanishes in the whole range $\rho \leq \rho_h$ and
    is small inside the horizon for $\gamma$ close to $\gamma_n(d)$.
    This allows us to find the critical values $\gamma_n(d)$ analytically:
    \eq ({\ref{e-phi}}) reduces to a linear equation for $f$ in a
    given (de Sitter) background and, combined with the boundary conditions
    $f(0) =0$ and $f(\rho_h) < \infty$, leads to a linear eigenvalue problem.
    Its solution (see \cite{BMP} for details) in the $d$-dimensional case
    gives the upper limits $\gamma_n (d) $ and the corresponding minimal
    horizon radii $r_{h}=r_{h n}$ for solutions with the Higgs field
    magnitude $f$ changing its sign $n$ times:
\bear
    r_{h n} \eql \sqrt{( 2n+1) (2n+d+2)/ \lambda \eta ^{2} },             \label{r h,n}
\\
    \gamma_n \eql \frac{2d(d+1) }{(2n+1)(2n+d+2)}.    \label{gamma n}
\ear
    For $d=2$ \eqs (\ref{r h,n}), (\ref{gamma n}) reduce to (52) in
    \Ref{BMP}. Under the condition (\ref{clas}) these solutions
    remain in the classical gravity regime.

\section{5-dimensional models and nonsingular cosmology}

\subsection {The extra dimension}

At present there is no evidence for the existence of more than
three spatial dimensions up to the achievable energies about
several hundred GeV. But this energy is quite tiny on the Planck
scale, $\sim 10^{19}$ GeV. Our solutions of possible cosmological
interest correspond to $\gamma > d-1$, i.e., the Planck energy
scale. Even under the condition (\ref{clas}), there remains an
enormous range of scales in the early Universe in which the number
of equally important spatial dimensions could have been greater
than 3.

If we try to consider our $d=3$ solutions in a cosmological
context, the extra coordinate is $t$ in (\ref {metric general})
and (\ref{metric}). The coordinate $t$ is time inside the horizon
and becomes a fourth spatial coordinate outside it, where $A(\rho)
<0.$ The metric (\ref{metric}) takes the form
\[
     ds^2 = \frac{d\rho^2}{|A(\rho) |}-| A(\rho) | dt^2-r^2(\rho)
    d\Omega_3^2.
\]
     Introducing the proper time $\tau $ of a co-moving observer
     outside the horizon,
\beq
     \tau = \int_{\rho_h}^{\rho} \frac{d\rho} {\sqrt{|A(\rho)|}},
    \label{proper time}
\eeq
     we arrive at a 5-dimensional Kantowski-Sachs cosmology with a
     closed Friedman-Robertson-Walker metric in the ($3+1$)-dimensional
     space-time section of constant $t$:
\beq
     ds_4^2 = d\tau^2 - a^2 (\tau) d\Omega_3^2
                - |A(\rho(\tau))| dt^2.       \label{FRW metric}
\eeq
     The 4-dimensional spherical radius $r(\rho) $ now plays
     the role of the scale factor: $a(\tau) =r(\rho (\tau))$.

     It is tempting to explain the unobservability of the extra dimension
     parametrized by the coordinate $t$ by compactifying $t$ with a certain
     ``period'' $T$ in the spirit of Kaluza-Klein models. Such a
     compactification would, however, lead to a singularity at $r=r_h$, as
     is clear from Fig.\,2. If $t\in \R$, the static region (the left
     quadrant of the diagram) is connected with the future cosmological
     region (the upper quadrant) by the horizon, crossed by photons as well
     as massive particles without problems. If, however, the $t$ axis is
     made compact by identifying, say, the points $t_1$ and $t_2$ on the $t$
     axis, then the static and cosmological regions in the diagram take the
     form of the dashed sectors, actually tubes of variable thickness,
     connected at one point only, the ends (tips) of the tubes. The
     curvature invariants do not change due to this identification and
     remain finite, and the emerging singularity in the $(\rho,t)$ plane
     resembles a conical singularity.

\begin{figure}\centering
    \includegraphics{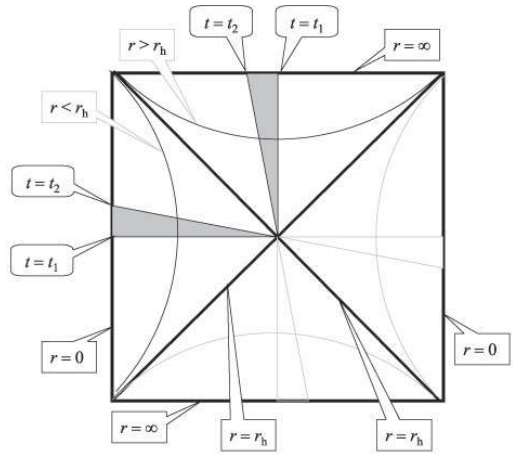}
    \caption{\label{fig:Figure2} Carter-Penrose diagram of a global
    monopole with a cosmological horizon. The diagonals of the square
    depict the horizons. After identification of $t_1$ and $t_2$, only
    the dashed regions survive.}
\end{figure}

    Compactification is not the only possibility of explaining why the $t$
    coordinate is invisible. Instead one can assume that, at some instant
    of the proper cosmological time $\tau$ of the 5-dimensional model
    (\ref{FRW metric}), a phase transition happens at a certain energy
    scale $1/T$ leading to localization of matter on the 3-spheres in
    the spirit of brane world models. Anyway, within our macroscopic theory
    without specifying the structure of the physical vacuum, it is impossible
    to explain why the extra dimension is not seen now. Nevertheless it is of
    interest to describe some cosmological characteristics of the $d=3$
    global monopole.

\subsection {Some cosmological estimates}

    For $d=3$ the internal integrals in (\ref{A(ro,rh)})
    and (\ref{ro h (gamma)}) have a logarithmic
    character, and instead of (\ref{r h at d>3}) and
    (\ref{A(r) at d>3})\ we get
\beq
    \gamma -2=\frac{3}{\lambda \eta
    ^2r_{h}^2}\left[ B+\ln \left( \lambda \eta
    ^2r_{h}^2\right) \right] ,\qquad r_{h}^2\gg \frac{1}{\lambda
    \eta ^2},\qquad d=3  \label{r h at d=3}
\eeq

\begin{figure}\centering
    \includegraphics{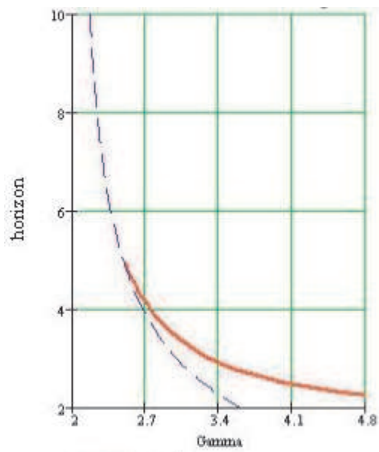}
    \caption{\label{fig:Figure3} The dimensionless horizon radius
    $\sqrt{\lambda}\eta r_{h}$ vs. $\gamma $\ for $d=3$ (solid line).
    The dashed line is the asymptotic dependence (\ref{r h at d=3}) valid for
    $\gamma-2 \ll 1.$}
\end{figure}

    and
\beq
     A(a) =-\frac{\gamma -2}{2\alpha^2}\left( 1-\frac{r_{h}^2}{a^2}\right)
     + \frac{3\gamma }{2\alpha ^2\lambda \eta^2} \frac{\ln (a/r_{h})}{a^2},
     \qquad    a>r_{h},\ \ d=3.     \label{A(r) outside rh}
\eeq The dependence $a(\tau) $ can be found from Eq.(\ref{proper
time}). In (\ref{r h at d=3}) $B$ is a constant about unity. Our
numerical estimate gives $B\approx 0.75$. The dimensionless radius
of the horizon $\sqrt{\lambda }\eta r_{h}$ as a function of
$\gamma$ is presented in \fig 3 for $d=3$ (solid line). The dashed
line is the asymptotic dependence (\ref{r h at d=3}) valid for
$\gamma - 2\ll 1.$ The function $A (\tau) \equiv A( a(\tau) ) $ is
shown in \fig 4 for $d=3$ and $\gamma = 3, 3.5$ and 4. The
numerical and analytical results are shown by solid and dashed
lines, respectively. It is remarkable that only for $\gamma = 4$
the approximate analytical dependence (\ref{A(r) outside rh}),
valid, strictly speaking, for $\gamma-2 \ll 1$, is slightly
different from the more precise one found numerically.


\begin{figure}\centering
    \includegraphics{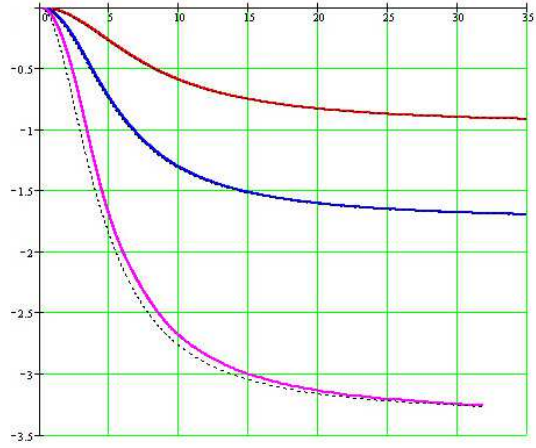}
    \caption{\label{fig:Figure4} The function $A( \tau) \equiv A( a( \tau)) $ for
    $d=3$ and $\gamma =$ 3, 3.5 and 4. Solid lines show numerical
    results while dashed lines the analytical dependence (\ref{A(r) outside rh}).}
\end{figure}

 Far outside the horizon $A(a) $ tends to a constant value:
\[
    A( a) \to -\frac{\gamma -2}{2\alpha ^2},\qquad a\gg r_{h},
\]
and the metric (\ref{FRW metric})\ describes a uniformly
expanding world with the linear dependence $a(\tau) $ at late times:
\beq
    a(\tau) =\alpha \sqrt{\left |A\left( \infty \right)
    \right| }\tau =\sqrt{\frac{\gamma -2}{2}}\tau ,\qquad \tau
    \to \infty .                         \label{a(Tau)}
\eeq

 The Hubble parameter $H = \dot a/a$, where the dot denotes $d/d\tau $,
 is found analytically from the expression (\ref{A(r) outside rh})
 for $A(a)$ [$d=3,\ a> r_h \gg 1/(\sqrt{\lambda }\eta)]$:
 \beq
    H(a) = \frac{1}{a}\sqrt{\frac{\gamma -2}{2}\left(
    1-\frac{r_{h}^2}{a^2}\right) -\frac{3\gamma }{2}
    \frac{\ln ( a/r_h) }{\lambda \eta^2 a^2}}.      \label{H(a)}
 \eeq


\begin{figure}\centering
    \includegraphics{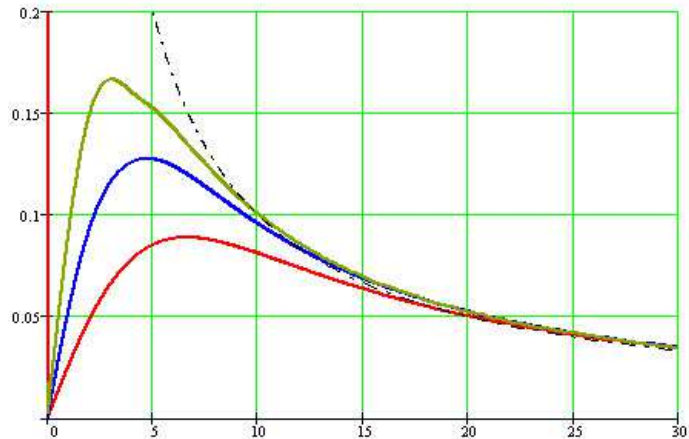}
    \caption{\label{fig:Figure5} The Hubble parameter $H(\tau)$
    for $\gamma =$ 3, 3.5, and 4. At late times
    $H(\tau)=1/\tau$ (dashed curve).}
\end{figure}

The temporal evolution of the Hubble parameter $H(\tau)$ is shown in \fig 5
for $\gamma =$ 3, 3.5, and 4. The expansion starts from the horizon at
$\tau = 0$ and rather quickly approaches the late-time behavior
$H(\tau) =\tau^{-1}$. We actually have the asymptotic regime almost
immediately after the beginning.

If we try to extrapolate this late-time regime to the present epoch, we can
use the estimate given in \Ref{MTW} (Box 27.4): $\dot a \approx 0.66$, and
\eqs (\ref{a(Tau)}) and (\ref{r h at d=3}) lead to
\beq
      \gamma= 2 + 2\dot{a}^2 = 2.87, \qquad  \sqrt{\lambda} \eta
                    r_{h}\approx 3.65.  \label{2.87)}
\eeq These estimates conform to the monopole parameter values
leading to a nonsingular cosmology.

 The symmetry-breaking potential (\ref{V(r)}), averaged over the
 oscillations, $V(\tau) \equiv \overline{V( a(\tau) )}$ is a
 decreasing function of $\tau$:
 \beq
    V(\tau)  =\frac{9}{(\gamma -2)^2\lambda \tau^4}
    + \frac{ \lambda \eta^4 C^2}
        {2[(\gamma/2 -1)\lambda \eta^2 \tau^2]^{3/2}},
           \qquad \tau \to \infty.              \label{V(Tau)}
 \eeq
 Scalar field potentials are often interpreted in cosmology as
 a time-dependent effective cosmological constant. A reason is that $V$
 enters as a $\Lambda$-term into the energy-momentum tensor. As is seen from
 (\ref{V(Tau)}), in our case this term behaves as a mixture of two
 components, one decaying with cosmological expansion like radiation
 ($\sim \tau^{-4} \sim a^{-4})$, the other like matter without pressure
 ($\sim \tau^{-3} \sim a^{-3})$ in 4 dimensions. The four-dimensional energy
 density corresponding to $V$ is proportional to $V\sqrt{|A|}$. However,
 at late times $\sqrt{|A|}$, the extra-dimension scale factor, tends to a
 constant. Hence the five- and
 four-dimensional behaviours of the energy density actually coincide at large $\tau$.
 One can say that the potential $V(\phi)$ in the global monopole model gives
 rise to both dark radiation and dark matter. Recall that, by modern views,
 both must necessarily be present in the Universe from the observational
 viewpoint \cite{Sahni and Starobinsky}.

 These estimates can only show that the 5-dimensional global monopole model
 is in principle able to give plausible cosmological parameters.
 Quantitative estimates certainly require a more complete model including
 further phase transitions, one of which should explain the unobservability
 of the fifth dimension.

\subsection*{Acknowledgement}

The authors are grateful to A.F. Andreev, A.A. Starobinsky, V.A.
Marchenko, and M. Yu. Kagan for useful discussions.

\bigskip

\bigskip

\bigskip

file.tex

\end{document}